\newcounter{myctr}
\def\myitem{\refstepcounter{myctr}\bibfont\noindent\ifnum\themyctr>9\else\phantom{0}\fi\hangindent17pt\themyctr.\enskip}

\documentclass{ws-ijqi}
\begin{document}
\markboth{Agung Trisetyarso and Rodney Van Meter}
{Circuit Design for A Measurement-Based Quantum Carry-Lookahead Adder}

\catchline{}{}{}{}{}

\title{CIRCUIT DESIGN FOR A MEASUREMENT-BASED\\ 
QUANTUM CARRY-LOOKAHEAD ADDER}

\author{AGUNG TRISETYARSO}

\address{ Department of Applied Physics and Physico-Informatics, Keio University,\\
3-14-1 Hiyoshi, Kohoku-ku, Yokohama-shi, Kanagawa-ken 223-8522, Japan\\
trisetyarso07@a8.keio.jp}

\author{RODNEY VAN METER}

\address{Faculty of Environment and Information Studies, Keio University,\\
5322 Endo, Fujisawa-shi, Kanagawa-ken 252-8520, Japan\\
rdv@sfc.wide.ad.jp}

\maketitle

\begin{history}
\received{Day Month Year}
\revised{Day Month Year}
\end{history}

\begin{abstract}
We present the design and evaluation of a quantum carry-lookahead adder (QCLA) using measurement-based quantum computation (MBQC), called MBQCLA. QCLA was originally designed for an abstract, concurrent architecture supporting long-distance communication, but most realistic architectures heavily constrain communication distances. The quantum carry-lookahead adder is faster than a quantum ripple-carry adder; QCLA has logarithmic depth while ripple adders have linear depth. MBQCLA utilizes MBQC's ability to transfer quantum states in unit time to accelerate addition. MBQCLA breaks the latency limit of addition circuits in nearest neighbor-only architectures : compared to the $\Theta(n)$ limit on circuit depth for linear nearest-neighbor architectures, it can reach $\Theta(log~n)$ depth. MBQCLA is an order of magnitude faster than a ripple-carry adder when adding registers longer than 100 qubits, but requires a cluster state that is an order of magnitude larger. The cluster state resources can be classified as computation and communication; for the unoptimized form, $\approx$ 88 $\%$ of the resources are used for communication. Hand optimization of horizontal communication costs results in a $\approx$ 12$\%$ reduction in spatial resources for the in-place MBQCLA circuit. For comparison, a graph state quantum carry-lookahead adder (GSQCLA) uses only $\approx$ 9 $\%$ of the spatial resources of the MBQCLA. 
\end{abstract}


\section[INTRODUCTION]{\label{sec:level1}Introduction}
Measurement-based quantum computation (MBQC) is a new paradigm for implementing quantum algorithms using a quantum cluster state \cite{Briegel:2009hs} \cite{Raussendorf01} \cite{Briegel01} \cite{Raussendorf03}. A cluster state is a highly entangled state of qubits which can serve as the resource for universal quantum computation. By subsequent single-qubit measurements, quantum gates are effected on the logical qubits encoded in the cluster state. Quantum information propagation in a cluster is driven by the pattern of measurement bases, regardless of the measurement outcomes. MBQC is attractive because cluster states are considered to be easy to create on systems ranging from the polarization state of photons \cite{Walther05} to Josephson junction qubits \cite{Nori01}.

A cluster state can be built on  a two-dimensional rectangular lattice with Manhattan geometry. Some of the qubits in the cluster state are data qubits, while the rest are created in a generic entangled state. Employing quantum correlations for quantum computation, quantum gates on the data qubits can be evaluated by measuring lattice qubits in a particular basis. All gates in the Clifford group, including CNOT, can be performed in one time step via a large number of concurrent measurements. Remarkably, because both wires and SWAP gates are in the Clifford group, MBQC supports long-distance gates in a single time step even when the cluster state is built on a  physical system permitting only nearest-neighbor interactions. \footnote{In this paper, we focus on the quantum rather than classical aspects of the system; a Pauli frame correction based on measurement results may be necessary and will be limited by classical signal propagation time. Thus, single time step wires depend on the assumption that classical signal propagation is fast compared to quantum measurements and gates.}  

The Toffoli Phase gate, which is not in the Clifford group, can be executed in two time steps, where the measurement basis for the second step is selected depending on previous measurement outcomes. This adaptive process, which must be cascaded through most interesting quantum circuits, determines the overall performance of many algorithms.

Thus, a cluster state can be used to execute arbitrary quantum algorithms. MBQC algorithms are often created by mapping known quantum circuits onto the cluster state. The challenge is to find application algorithms that match the strengths of MBQC. Here, we choose to address the problem of integer addition.

Addition is a critical subroutine for algorithms such as Shor's algorithm for factoring large numbers \cite{Sho97SIComp} \cite{Vedral96} \cite{Beckman01} \cite{VanMeter1}. Addition can be executed in many ways, with its performance being primarily dependent on carry propagation, which is normally limited by the physical architecture\cite{MeterOskin} \cite{Byung01}. The simplest method is ripple-carry addition, which has depth of $\Theta(n)$\cite{Vedral96} \cite{Beckman01} \cite{cuccaro-2004} \cite{Takahashi01} to add two $n$-bit numbers. In a ripple-carry adder, carry information is propagated from the low-order qubits to the high order qubits one step at a time.   

The goal of our work is to reduce the execution time of addition on MBQC. Raussendorf \textit{et al.} successfully mapped the VBE ripple-carry adder to MBQC bending the circuit layout to reduce the spatial resources\cite{Raussendorf03} \cite{Vedral96}. However, a ripple-carry adder does not take good advantage of the strengths of MBQC. By unifying the quantum carry-lookahead adder (QCLA)\cite{Draper06} with MBQC, we have designed a much faster circuit for large $n$. In this paper, we present our design for the MBQCLA and evaluate the design in terms of its execution speed and resource requirements. The depth and spatial optimizations are also discussed.

The paper is organized as follows: the basic notions of measurement-based quantum computation and quantum carry-lookahead adder are given in Section 2. Section 3 contains the implementation of MBQC form for QCLA circuits. Here, the out-of-place, in-place, and the optimized version of in-place circuits are discussed to obtain their performances and requirements. We conclude in Section 4. Appendix A gives a detailed exposition of a NOT gate in cluster state and the graphical notation used in this paper. Appendix B contains the procedures to implement the out-of-place and in-place QCLAs in abstract quantum circuit form. Appendix C provides the requirements and performance for MBQCLA circuits.

\section[Background]{\label{sec:level2}Background}
\noindent

Our proposed circuits build on two concepts: (a) measurement-based quantum computation, and
(b) the quantum carry-lookahead adder. In this section, we will present a short review of these concepts.

\subsection[Measurement-Based Quantum Computation]{\label{sec:level2a}Measurement-Based Quantum Computation}
\noindent
\\A one-dimensional cluster state is in the form of \begin{equation}\left|\Phi_{\textit{N}}\right\rangle =\frac{1}{2^{\frac{\it{N}}{2}}}\bigotimes^{\textit{N}}_{\textit{a}=1}\left(\left|0\right\rangle_{\textit{a}}\sigma^{(\textit{a}+1)}_{\upsilon}+\left|1\right\rangle_{\textit{a}}\right),\end{equation} where $\sigma$$_{\upsilon}$$^{(i)}$ is the Pauli operator operating on qubit $\it{i}$ in the $N$-qubit cluster ${\mathcal C}$, $\textit{a}$ is the index of a qubit in cluster ${\mathcal C}$ and $\upsilon$ can be $\textit{x,~y,}$ or $\textit{z}$ depending on the choice of interaction Hamiltonian between neighbors \cite{Walther05} and with the convention $\sigma^{\textit{N}+1}_{\upsilon}=1$. In general, the cluster state should obey the quantum correlation equation\begin{equation} \bigotimes_{a,\upsilon}~\sigma^{(\textit{a})}_{\upsilon}\left|\Phi_{\left\{\textit{k}\right\}}\right\rangle_{{\mathcal C}}=\left(-1\right)^{\textit{k}_{\textit{a}}}\left|\Phi_{\left\{\textit{k}\right\}}\right\rangle_{{\mathcal C}}~,\end{equation}
where $\upsilon$=$\textit{I,~x,~y,~z}$, and $\textit{k}\textit{$_{a}$}=\left\{0,~1\right\}$. The parameter $\{$$\textit{k}$$\}$ is a set of index parameters specifying the cluster state. $\left|\Phi_{\left\{\textit{k}\right\}}\right\rangle_{\mathcal C}$ expresses the cluster state before the measurement and $\left|\Psi_{\left\{\textit{k}\right\}}\right\rangle_{\mathcal C (\it{g})}$ represents the cluster state after a set of measurements in which quantum gate $g$ has been simulated on the cluster state. 

The cluster state can be created in several ways, e.g., initializing every qubit to the $|$+$\rangle$ state and performing a Controlled-Z gate between each neighboring pair. With such a cluster state, Raussendorf \textit{et al.} showed that a carefully chosen measurement pattern can effect any quantum gate on logical qubits.

Suppose we have an initial set of cluster state eigenvalue equations, $\left|\Phi_{\left\{\textit{k}\right\}}\right\rangle_{{\mathcal C}}$, representing the cluster which is the union of the input cluster ($\mathcal C$$_{input}$), the machine cluster, (${\mathcal C}$$_{machine}$) and the output cluster (${\mathcal C}$$_{output}$). All of the qubits except the output cluster are measured by the projective measurement operators ${\mathcal P}$ with certain measurement patterns ${\mathcal M}$, ${\mathcal P}_{\{s\}}$$^{({\mathcal C})}$(${\mathcal M}$)=$\bigotimes_{\scriptsize{k \in {\mathcal C}}}$$\frac{1+(-1)^{k_{a}}\vec{r}_{k}.\vec{\sigma}^{(k)}}{2}$. The new $m$-qubits output register of the quantum logic network is the cluster state $\left|\Psi_{\left\{\textit{k}\right\}}\right\rangle_{\mathcal C(g)}$ obeying 2$m$ new eigenvalue equations:

\begin{equation}
\sigma^{({\mathcal C}_{input}(g),\it{i})}_{\textit{x}}(\textit{U}\sigma^{(\textit{i})}_{\textit{x}}\textit{U}^{\dagger})^{({\mathcal C}_{output}(g))}\left|\psi\right\rangle_{{\mathcal C}(g)}=(-1)^{\lambda_{\textit{x,i}}}\left|\psi\right\rangle_{{\mathcal C}(g)}
\end{equation}  

\begin{equation}
\sigma^{({\mathcal C}_{input}(g)\it{i})}_{\textit{z}}(\textit{U}\sigma^{(\textit{i})}_{\textit{z}}\textit{U}^{\dagger})^{({\mathcal C}_{output}(g))}\left|\psi\right\rangle_{{\mathcal C}(g)}=(-1)^{\lambda_{\textit{z,i}}}\left|\psi\right\rangle_{{\mathcal C}(g)},
\end{equation} 
\noindent
where $\lambda_{\textit{x,i}}$, $\lambda_{\textit{z,i}}$ $\in$ $\{0, 1\}$ are the measurement outcomes.
\indent
\\
Quantum computation in Clifford algebra form implicitly appears in the final set of eigenvalue equations after the measurements. A brief review is given in Appendix A.

Several remarkable properties follow \cite{Raussendorf01} \cite{Browne01}: 

\begin{itemlist}
\item Measurement of qubits in the machine cluster in the $\sigma_{z}$-eigenbasis removes them from the main cluster and \textbf{\textit{disconnects}} all of the their bonds.
\item Measurement of qubits in the machine cluster in the $\sigma_{x}$-eigenbasis removes them from the main cluster and creates \textbf{\textit{Bell pairs}} between the qubits in input cluster and the qubits in the output cluster.
\item Measurement of qubits in the machine cluster in the $\sigma_{y}$-eigenbasis removes them from the main cluster and leaves an entangled state between the qubits in the input cluster and the qubits in the output cluster.
\end{itemlist}

The measurement calculus is a convenient formalism for representing MBQC quantum gates \cite{Vanos1} \cite{Voufo1}. Danos \textit{et al.} show how to write an MBQC quantum gate ${\mathcal U}$ in the form ${\mathcal U}$:=($\{$Resources$\}$$_{\mathcal U}$, $\{$Input$\}$$_{\mathcal U}$, $\{$Output$\}$$_{\mathcal U}$, $\{$$EMC$$\}$$_{\mathcal U}$). Based on this definition, we introduce the notation ${\mathcal U}^{<n>}$ meaning a quantum gate in MBQC using $n$ qubits. ${\mathcal {CNOT}}^{<4>}$ refers to ${\mathcal {CNOT}}^{<4>}$ := ($\{$1, 2, 3, 4$\}$, $\{$1, 4$\}$, $\{$3, 4$\}$, $\{$$X_{4}^{s_{3}}Z_{4}^{s_{2}}Z_{1}^{s_{2}}M_{3}^{x}M_{2}^{x}E_{13}E_{23}E_{34}$$\}$). A fifteen-qubit form of the gates is :${\mathcal {CNOT}}^{<15>}$=($\{$Resources$\}$$_{{\mathcal {CNOT}}^{<15>}}$, $\{$Input$\}$$_{{\mathcal {CNOT}}^{<15>}}$, $\{$Output$\}$$_{{\mathcal {CNOT}}^{<15>}}$, $\{$$EMC$$\}$$_{{\mathcal {CNOT}}^{<15>}}$) := ($\{$1, ... , 15$\}$, $\{$1, 9$\}$, $\{$7, 15$\}$, $\{$$EMC$$\}$$_{{\mathcal {CNOT}}^{<15>}}$) as mentioned in Ref.~\refcite{Raussendorf03}. There are two types of Toffoli gate: ${\mathcal {CCNOT}}^{<54>}$  \cite{Raussendorf03} and  ${\mathcal {CCNOT}}^{<39>}$ \cite{Voufo1}. Both Toffoli gates have similar numbers of adaptive measurements, but different numbers of qubit resources. ${\mathcal {CCNOT}}^{<39>}$ must be connected into an arbitrary graph, while  ${\mathcal {CCNOT}}^{<54>}$ is appropriate for the Manhattan geometry cluster state. 

The physical implementation of MBQC requires a lattice system with an Ising-like interaction between the qubits so that the quantum information can be propagated in the lattice due to the measurement. Several physical implementations have been proposed; Meier \textit{et al.} proposed the possibility of experimental realization to perform initialization, quantum gate operation and read-out mechanism in antiferromagnetic spin cluster quantum computing\cite{Meier02}. Devitt \textit{et al.} have described an all-optical implementation where the required number of photonic modules and chips only depends on the cross section length of the two-dimensional lattice (corresponding to the $y$-axis in our figures)\cite{Politi01} \cite{Devitt02}.  

MBQC runs in two phases: prepare the cluster state, then measure. Because the preparation step is completely generic, failure in coupling the qubits is not a problem. Mechanisms that succeed only probabilistically can be used, as long as failures are heralded, making optical QC suitable for MBQC.\cite{louis2007efficiencies}

\subsection[Quantum Carry-Lookahead Adder]{\label{sec:level2b}Quantum Carry-Lookahead Adder}

The Quantum Carry-Lookahead Adder (QCLA) was designed by Draper \textit{et al.}\cite{Draper06}. The quantum carry-lookahead adder is potentially more efficient than a quantum ripple-carry adder since its depth is $\Theta(log~(n))$. A carry-lookahead adder uses three phases, the ``Generate'' (G), ``Propagate'' (P), and ``Kill'' (K) networks, each of which progressively doubles the length of its span in each time step, to calculate the complete ``Carry'' values (C). In practice, the networks are somewhat redundant, and Draper \textit{et al.} defined their circuit using only the P and G networks to calculate the final carry C. The out-of-place form of the QCLA performs the unitary transformation $\left|\it{a,b,0}\right\rangle \longrightarrow \left|\it{a,b,a}+\it{b}\right\rangle$, and the in-place form calculates $\left|\it{a,b}\right\rangle \longrightarrow \left|a,\it{a}+\it{b}\right\rangle$ where $\left|\it{a}\right\rangle, \left|\it{b}\right\rangle$ and $\left|\it{a}+\it{b}\right\rangle$ are $\it{n}$-qubit registers, where \textit{0} is the low-order qubit and \textit{n-1} is the high-order qubit.   

The carry-lookahead adder starts with an initial addition round, consisting of a half adder for each qubit in the logical register. Starting from the basic idea of the carry-lookahead adder originally designed for classical binary logic\cite{Ercegovac}, the carry is propagated from bit to bit $i$$\rightarrow$$j$$\rightarrow$$k$, where $i$$\leq$$j$$\leq$$k$, so carry equations or majority blocks are represented by:
\begin{equation}
c_{\it{j}}=g[\it{i},\it{j}]\oplus p[\it{i},\it{j}]\wedge c_{\it{i}}
\end{equation}
\begin{equation}
c_{\it{k}}=g[\it{j},\it{k}]\oplus p[\it{j},\it{k}]\wedge c_{\it{j}}
\end{equation}

By straightforward substitution of these equations, we have the equation:
\begin{equation}
c_{\it{k}}=g[\it{j},\it{k}]\oplus p[\it{j},\it{k}]\wedge (g[\it{i},\it{j}]\oplus p[\it{i},\it{j}]\wedge c_{\it{i}})
\end{equation}
or
\begin{equation}
\label{eq8}
c_{\it{k}}=g[\it{j},\it{k}]\oplus p[\it{j},\it{k}]\wedge g[\it{i},\it{j}]\oplus p[\it{i},\it{j}]\wedge p[\it{j},\it{k}]\wedge c_{\it{i}}
\end{equation}

Substituting by c$_{\it{k}}$=g[$\it{i},\it{k}$]$\oplus$ p[$\it{i},\it{k}$]$\wedge$ c$_{\it{i}}$ into Eq.(\ref{eq8}) gives:

\begin{equation}
g[\it{i},\it{k}]=g[\it{j},\it{k}]\oplus p[\it{j},\it{k}]\wedge g[\it{i},\it{j}].
\end{equation}
A circuit that performs this computation in a lookahead adder is called the Generate network. Similarly, a circuit that implements the equations
\begin{equation}
\textit{p}[\textit{i},\textit{k}]=\textit{p}[\textit{i},\textit{j}]\wedge \textit{p}[\textit{j},\textit{k}]
\end{equation}
for any $i$$<$$j$$<$$k$ is called the Propagate network. The implementation of these networks in reversible computation is realized by the following steps where $n$ is logical qubits, $t$ is the round number and $m$ is the index of qubits in the register:
\begin{enumerate}
\item $P$-rounds. For $t$=1 to $\lfloor$log $n$$\rfloor$ - 1: for 1$\leq$$m$$<$$\lfloor$$n$/2$^{t}$$\rfloor$. Then the connection between the steps in this round is expressed by: $P_{t}[m]$$\oplus$=$P_{t-1}$[2$m$]$P_{t-1}$[2$m$+1].
\item $G$-rounds. For $t$=1 to $\lfloor$log $n$$\rfloor$: for 0$\leq$$m$$<$$\lfloor$$n$/2$^{t}$$\rfloor$. The relation between the steps in the round is: $G$[2$^{t}$$m$+2$^{t}$]$\oplus$=$G$[2$^{t}$$m$+2$^{t-1}$]$P_{t-1}$[2$m$+1].
\item $C$-rounds. For $t$=$\lfloor$log 2$n$/3$\rfloor$ down to 1: for 1$\leq$$m$$<$$\lfloor$($n$-2$^{t-1}$)/2$^{t}$$\rfloor$. The connection between the steps in the round is represented by: $G$[2$^{t}$$m$+2$^{t-1}$]$\oplus$=$G$[2$^{t}$$m$]$P_{t-1}$[2$m$].
\end{enumerate}

These networks will be applied both to out-of-place and in-place QCLA circuits. Those circuits, which are distinguished by the form of the addition scheme, are explained in more detail in Appendix B. 

\subsection[Space and Depth Complexities Trade-off in Nearest-Neighbor Interactions Architectures]{\label{sec:level3p}MBQC as Solution for Long-Distance Communication in Nearest-Neighbor Architectures}

The QCLA circuit explained above is one example of a circuit design that assumes communication between non-adjacent qubits is allowed. However, scalable quantum computers may allow only nearest-neighbor interactions\cite{Lloyd01}. The depth complexity of a circuit on a Nearest-Neighbor (NN) architecture may be larger than non-NN architectures. Under some circumstances, MBQC gives us trade off between depth and space complexity\cite{broadbent01}: one can reduce the circuit depth by adding a number of measurements, entanglements, and byproduct operations in the quantum circuit. 

The out-of-place and in-place QCLA respectively have the overall depth $\lfloor$$log_{2}$($\it{n}$)$\rfloor$+$\lfloor$$log_{2}$($\it{n}$/3)$\rfloor$+4 and $\lfloor$$log_{2}$($\it{n}$)$\rfloor$+$\lfloor$log($n$-1)$\rfloor$+$\lfloor$$log_{2}$($\it{n}$/3)$\rfloor$+$\lfloor$$log_{2}$$\frac{n-1}{3}$$\rfloor$+8. However, these abstract quantum circuits assumed unrealistic conditions: interactions between non-adjacent qubits can be perfectly implemented. When application qubits one assigned positions in a quantum computer, some qubits we wish to interact may be widely separated; examining the circuit diagram for QCLA shows many long-distance gates crossing over many other qubits. In a nearest-neighbor architecture, we must swap qubits, step by step, until our desired qubits become neighbors. On a single line, the $\Theta(log_{2}(n))$ time steps for QCLA expands to $\Theta(n)$.\cite{Byung01}

\section[Quantum Carry-Lookahead Adder for Measurement-Based Quantum Computation]{\label{sec:level3}Quantum Carry-Lookahead Adder for Measurement-Based Quantum Computation}

This section explains the implementation of QCLA for MBQC. The performance and requirements for both out-of-place and in-place MBQCLA circuits schemes are evaluated. First we describe our metrics for evaluating circuits. The exposition of direct mapping on both schemes is given, followed by the optimization for the in-place circuit. The optimization is done by adjusting the border between the rounds of the circuit, then removing the unnecessary lattice sites between the quantum gates, reducing the communication costs in the circuit. For comparison, a graph state form of QCLA is also presented. More detailed results are presented in Appendix C.

\subsection[Evaluating Algorithms Executed using MBQC]{\label{sec:level3a}Evaluating Algorithms Executed using MBQC}

Logical quantum circuits can be evaluated based on the execution time, or \textit{circuit depth} (usually measured in numbers of Toffoli gates), number of qubits used, and total number of gates executed. The number of logical qubits is the number of required input qubits plus the number of ancillae required in the circuit representation of the algorithm.

We propose the use of  (a) the number of qubits in the cluster state,  (b) the number of clustering operations, (c) circuit area, and (d) circuit depth as measures of performance and cost for algorithms executed using MBQC. The number of cluster operations is the number of successful interactions needed. The circuit area is the height of the cluster times its width, assuming a regular rectangular lattice. All of these measures can be expressed in terms of problem size; in our case, in terms of $n$, the length of each of the logical registers being added.

\begin{figure} [!ht]
\centerline{\psfig{file=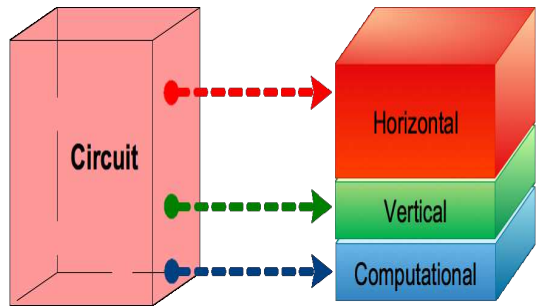, width=70mm, height=40mm}} 
\vspace*{13pt}
\caption{\label{compr1}Illustration for circuit costs for MBQCLA. On a two-dimensional Manhattan grid, the costs contain computational and communication costs. The resources for communication costs can be separated into two types of resources: horizontal (wires) and vertical (SWAP gates).}
\end{figure}

The goal of this paper is to minimize the execution time (d), while the other three (a-c) are measures of the cost. These costs, as shown in Figure \ref{compr1}, can be divided into two categories: $\it{first}$, $\textbf{\textit{computational resources}}$, i.e., the number of cluster qubits required for Toffoli Phase, CNOT and NOT gates. $\it{Second}$, $\textbf{\textit{communication resources}}$, i.e., the number of cluster qubits required for SWAP gates and wires. A circuit which uses no communication resources is called an $\textbf{optimal circuit}$. As noted above, MBQC requires a measure-adapt-measure cycle to implement non-Clifford gates. The execution time is the number of rounds of measurement, followed by computation of the adaptive bases for the next round.

In general, the optimal circuit resources can be determined by summing the resources consumed by the various types of computational gates. Thus, it  is expressed by
\begin{equation}
\label{opteqgen}
\sum_{i~\in~Quantum~Gate}{\mathcal X_{i}}{\mathcal R_{i}}
\end{equation}  
\noindent
where
\begin{itemlist}
\item ${\mathcal X_{i}}$ = Number of quantum gates of type $i$\\
\item ${\mathcal R_{i}}$ = Qubit resources for an $i$ gate 
\end{itemlist}

Because the QCLA is structured in a set of rounds, each of which contains only gates of a single type (e.g. Toffoli Phase Gate), we can discuss the cost in those terms. The cost of a null circuit would simply be the number of logical qubits multiplied by the cost of a horizontal wire. In an actual circuit, we replace some sections of horizontal wire with logical gates, and add vertical wires with SWAP gates as necessary to implement the logic. Thus, each round in the QCLA, when mapped onto the cluster state, is as wide as necessary to accommodate the necessary gate type.\footnote{Usually the width of a Toffoli gate.}

By considering the number of SWAP, Toffoli and CNOT gates in the initial addition circuit and in P,G and C networks as shown above, we can approximate the physical resources needed for an $n$-qubit out-of-place MBQCLA following this expression:
\begin{equation}
\label{sizeequation}
\textit{Size} \approx\sum_{\it {i}}({{\mathcal V}(n)}\times {{\mathcal B}_{i}}(n)\times {{\mathcal T}_{i}}+{{\mathcal X}_{i}}(n)~({{\mathcal R}_{i}}-{{\mathcal E}_{i}}\times {{\mathcal T}_{i}}))+ {{\mathcal R}_{SWAP}}\times {{\mathcal S}_{SWAP}}
\end{equation}
where $i~\epsilon$ $\{$Toffoli Phase, CNOT, and NOT Gates$\}$, ${\mathcal V}$$(n)$~= number of logical qubits, ${\mathcal B}$$_{i}$($n$)~= number of rounds for the gate of type $i$, ${\mathcal X}$$_{i}$($n$)~= number of gates of type $i$,  ${\mathcal T}$$_{i}$~= width of the $i$-gate (in lattice sites), ${\mathcal R}$$_{i}$ = number of lattice qubits in an $i$-gate, ${\mathcal E}$$_{i}$~= number of logical qubits in an $i$-gate (generally, one to three), ${\mathcal R}$$_{SWAP}$ is number of lattice qubits in a cluster state SWAP gate and ${\mathcal S}$$_{SWAP}$~= number of SWAP gates in a QCLA circuit, which is dependent on the mapping of the logical qubits to positions on the lattice.          

In Equation (\ref{sizeequation}), $\sum_{i}({\mathcal V}(n) \times {\mathcal B}_{i} \times {\mathcal T}_{i})$ is the cost of a lattice large enough to hold all of the circuit rounds that use type $i$-gates (that is horizontal communication costs). $-\sum_{\it {i}}{\mathcal E}_{i}\times {\mathcal X}_{i} \times{\mathcal T}_{i}$ is an adjustment for replacing wires with logic gates. ${\mathcal R}_{SWAP} \times {\mathcal S}_{SWAP}$, which depends on type of rounds in QCLA circuit, is the vertical communication in the circuit, which is entirely SWAP gates resources.   

Proposed implementations of cluster state quantum computing in solid-state technologies, which need Ising-like Hamiltonian\cite{Nori01} \cite{Nori02}
\cite{Levy01} \cite{Levy02} \cite{Meier02}, operate on a fixed 2-D lattice. Hence, they require wires for communication between the rounds, which means those proposals will use our optimized circuit. A photonic-based quantum computer\cite{Devitt02}\cite{Politi01} will require no wires for communication between the rounds, allowing the optimal circuits or graph state form to be used more or less directly. 

\subsection[Out-of-place Measurement-Based Quantum Carry-Lookahead Adder]{\label{sec:level3b}Out-of-place Measurement-Based Quantum Carry-Lookahead Adder}

\begin{figure} [!htp]
\centerline{\psfig{file=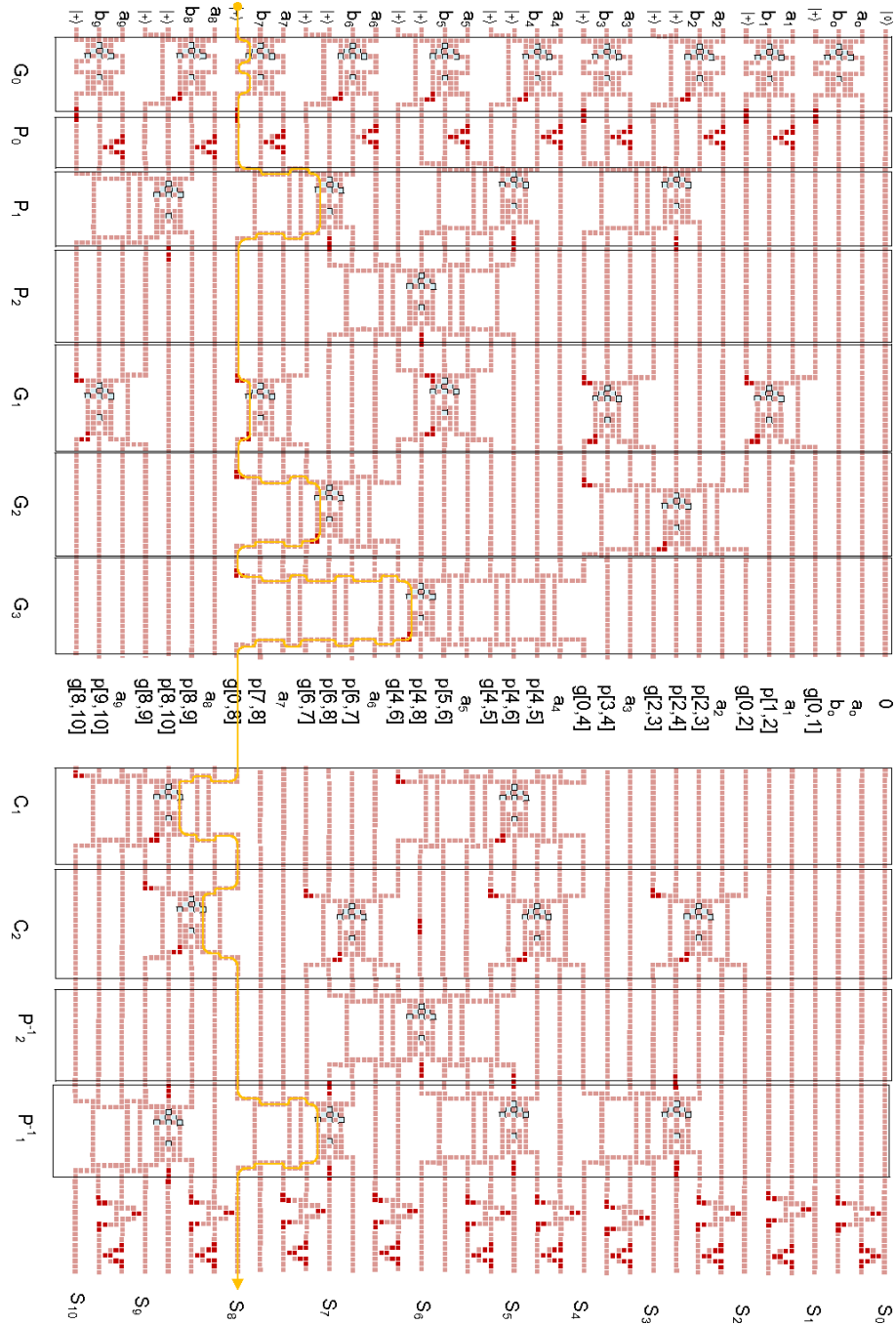, width=120mm, height=180mm}} 
\vspace*{13pt}
\caption{\label{out-of-placembqcqcla}Out-of-place MBQCLA. For $n$=10, the circuit consists of: 4 addition blocks, 9 rounds of gates for the carry networks (2 Propagate, 3 Generate, 2 Inverse Propagate and 2 Carry networks). For explanation of the colors, see Appendix A.}
\end{figure}

Our design for a 10-bit form of the out-of-place QCLA on MBQC is shown in Figure \ref{out-of-placembqcqcla}. The input qubits are on the left (top in the rotated figure) and output states are on the right. In the figures in this paper, a pink square qubit represents a cluster qubit measured in the $\sigma_{x}$-eigenbasis, a green qubit is for $\pi$-rotation on $\sigma_{x}$-eigenbasis measurement, a red qubit is for $\sigma_{y}$-eigenbasis measurement and a blue qubit is for $\sigma_{z}$-eigenbasis measurement. The propagation pattern of one logical qubit is highlighted in yellow. Our logical qubits are spaced with a pitch of four lattice sites to accommodate the necessary spacing between gates. Each large box outlines one round in the P, G or C networks. The circuit is presented in unoptimized form for clarity. 

This circuit is essentially a direct mapping of the abstract out-of-place QCLA (Figure \ref{abstractqcla} in Appendix B) to MBQC. The logical gates used are those described in Figure \ref{mbqcquantumgates} in Appendix A.  As noted above, in addition to the computational resources, we must add wires and SWAP gates. The long distance gates from the abstract out-of-place QCLA (Figure \ref{abstractqcla}) are executed using the scheme for non-adjacent computation (Figure \ref{Nonadjacentcomputation}). To completely characterize the circuit, we need to know how many SWAP gates and wire segments are added to complete the circuit. The exact cost depends on the layout of logical qubits. Below we calculate the number of SWAP gates required assuming the data layout of Figure \ref{out-of-placembqcqcla}. 

The abstract circuit consists of addition and carry computation circuits. For adding two $n$-qubit registers, the addition circuit is built from $\it{n}$ Toffoli gates and 3$\it{n}$-1 CNOT gates while the carry computation machinery consists of 4$\it{n}$-3$w(n)$-3$\lfloor log_{2}(n)\rfloor$-1 Toffoli gates. The number of Toffoli gates in this circuit can be obtained by adding the number of Toffolis in the addition, P, G, and C networks. For the out-of-place QCLA circuit, we have $\it{n}$-$w(n)$-$\lfloor log_{2}(n)\rfloor$ Toffoli gates for the P network, $\it{n}$-$w(n)$ Toffoli gates for the G network and $\it{n}$-$\lfloor log_{2}(n)\rfloor$-1 for C network, where $w(n)$ is the Hamming weight of the binary representation of $n$. Furthermore, the number of SWAP gates, which is the vertical communication resources, can be obtained as follows:

\begin{itemlist}
	\item The initial addition round needs 
\begin{equation}
\label{SWfAdd}
{\mathcal S_{{\mathcal Ad}}}=4n-2w(n)-2\lfloor log_{2}(n)\rfloor
\end{equation} 
\noindent
SWAP gates. For $\it{n}$=10, we need 30 SWAP gates consuming 360 lattice qubits. 
\end{itemlist}

\begin{itemlist}
	\item The propagate network needs 28 SWAP gates for $n$=10. This number can be obtained from the number of Toffoli gates for each round, $\lfloor$$n$/2$^{t_{p}}$$\rfloor$-1 where $t_{p}$ is the round number in the propagate network, 1 $\leq$ $t_{p}$ $\leq$ $\lfloor log_{2}(n) \rfloor$-1 .There are 4 Toffoli gates in the first P round with 16 SWAP gates and 1 Toffoli gate in the second P round with 12 SWAP gates. The vertical communication in P networks is
\begin{equation}
\label{SWfP}
{\mathcal S_{{\mathcal P}}}=\sum _{t_{p}=1}^{log_{2}(n)-1} 2n-2^{t_{p}+1}
\end{equation}	
\end{itemlist}
    
\begin{itemlist}
	\item Similarly, the generate network requires 58 SWAP gates for $n$=10. The number of Toffoli gates for each round is 5$n$+$w(n)$+2$\lfloor$$log_{2}(n)$$\rfloor$. As formulated by $\lfloor$$n$/2$^{t_{g}}$$\rfloor$, we have three rounds of generate networks; the first round consists of 5 Toffoli gates with 12 SWAP gates, the second round needs 2 Toffoli gates with 20 SWAP gates and the third round requires 1 Toffoli gate with 26 SWAP gates. The SWAP gates resources in the G network can be approximated by
\begin{equation}
\label{SWfG}
{\mathcal S_{{\mathcal G}}}=\sum _{t_{g}=1}^{log_{2}(n)} 4\lfloor log_{2}(n)\rfloor+2^{tg+1}\lfloor log_{2}(t_{g})\rfloor
\end{equation}
\noindent
where $t_{g}$ is the round in the G network.
\end{itemlist}

\begin{itemlist}
	\item The number of Toffoli gates in the carry network is $\lfloor \frac{n-2^{t_{c}-1}}{2^{t_{c}}} \rfloor$. Therein, the first Carry round has 4 Toffoli gates with 16 SWAP gates  and its second round has 22 SWAP gates. The resources of SWAP gates in the C network is
\begin{equation}
\label{SWfC}
{\mathcal S_{{\mathcal C}}}=\sum _{t_{c}=1}^{\lfloor log_{2}\frac{2n}{3}\rfloor} 2n-2\lfloor log_{2}(t_{c})\rfloor
\end{equation}
\noindent
where $t_{c}$ is the round number in the C network.
\end{itemlist}

\noindent
Following Equation (\ref{sizeequation}), ${\mathcal S}_{SWAP}$ is the sum of Equations (\ref{SWfAdd}), (\ref{SWfP}), (\ref{SWfG}), and (\ref{SWfC})
\begin{equation}
\label{oopeq}
{\mathcal S}_{SWAP}={\mathcal S_{{\mathcal Ad}}} + {\mathcal S_{{\mathcal P}}} + {\mathcal S_{{\mathcal G}}} + {\mathcal S_{{\mathcal C}}}.
\end{equation}

\begin{figure*}[!htp]
\begin{center}
{
    \label{fig:sub:a}
    \includegraphics[height=48mm]{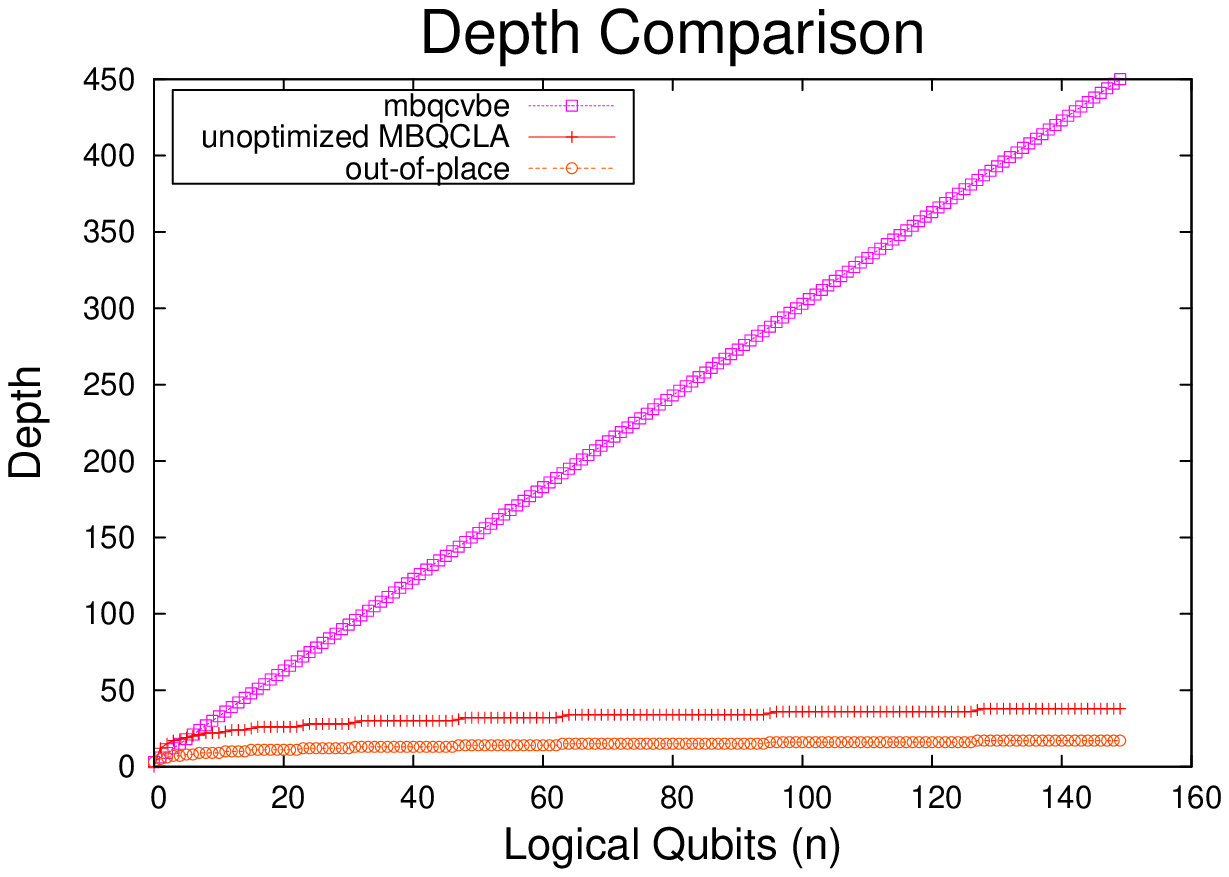}
}
\hfil
{
    \label{fig:sub:b}
    \includegraphics[height=48mm]{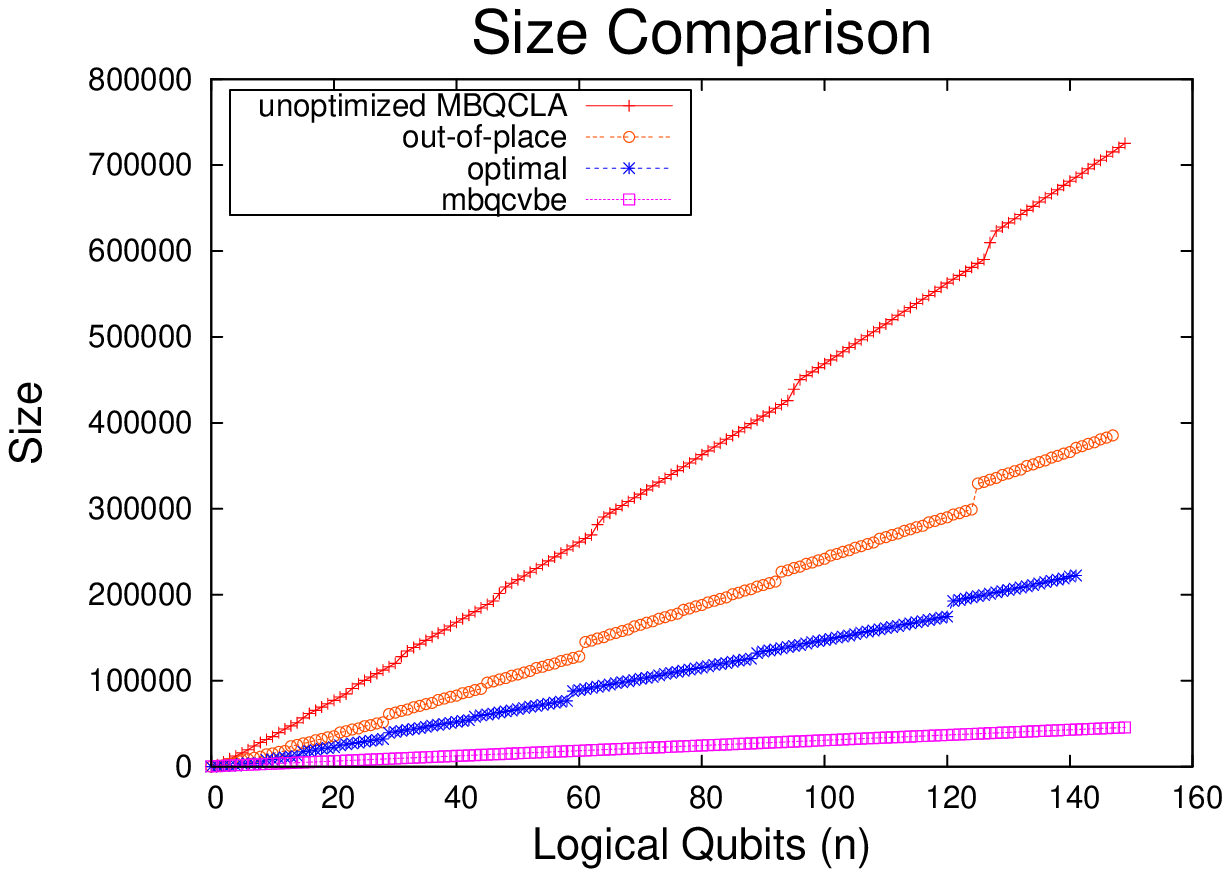}
}
\label{fig:sub} 
\vspace*{0.225truein}
\caption{\label{comparison1}Size and depth comparison between MBQC VBE and MBQCLA.``+'',``$\star$'', ``$\circ$'' and ``$\square$'' marks are for in-place, optimized in-place, out-of-place and MBQC VBE circuits, respectively.}
\end{center}
\end{figure*}

For the out-of-place MBQCLA, we see that the depth is reduced to $\lfloor$log$_{2}$($\it{n}$)$\rfloor$+$\lfloor$log$_{2}$($\it{n}$$/$3)$\rfloor$
+7 compared to $\approx$3$n$ for the VBE ripple-carry. However, this circuit costs more in physical resources, $\approx$901$\it{n}$+224$\it{n}$$\times$$\lfloor$log$_{2}$($\it{n}$)$\rfloor$ compared to $\approx$304$\it{n}$ for the VBE ripple-carry. The comparison of size and depth between MBQC VBE and MBQCLA is shown in Figure \ref{comparison1}.

\subsection[In-place Measurement-Based Quantum Carry-Lookahead Adder]{\label{sec:level3c}In-place Measurement-Based Quantum Carry-Lookahead Adder}

The next step is obtaining the performance and requirements of the in-place quantum carry look-ahead adder. Following the scheme in Ref.~\refcite{Draper06}, the erasure (uncomputation) of the low-order $n$-1 bits of the carry string $c$ requires additional circuitry. The algorithm for the in-place form is more complex than out-of-place and uses about twice as many Toffoli gates.

The subsequent procedures, as provided in Appendix B, give in-place MBQCLA circuit horizontal resources, as summarized in Table \ref{table2} (Appendix C). As shown in the previous section, the vertical communication resources can be estimated by counting the number of SWAP gates in the circuit. Because the in-place circuit uses more ancillae, which we interleave with the other qubits, the number of SWAP gates for the initial round of half-adders increases to 6$n$-4$w(n)$-4$\lfloor$log$n$$\rfloor$-2. The other SWAP gate resources can be obtained by examining the Propagate, Generate, and Carry networks. In the in-place circuit we need both for computing and uncomputing the carry status, meaning 8$\it{n}$+14$\lfloor$log $\it{n}$$\rfloor$+2 SWAP gates are needed for non-adjacent quantum computation. Straightforwardly, the physical resources for in-place circuit are
\begin{equation}
\label{ipsize}
\approx~2896n+64n\lfloor log_{2}(n)\rfloor.  
\end{equation}
\noindent
Also, by the use of Equations (\ref{SWfAdd})(\ref{SWfP})(\ref{SWfG}) and (\ref{SWfC}) for the out-of place circuit, the vertical communication resources (SWAP gates), or ${\mathcal S}_{SWAP}$ for in-place MBQCLA circuit is
\begin{equation}
\label{optinpl}
{\mathcal S}_{SWAP}={\mathcal S_{{\mathcal Ad}}} + {\mathcal S^{\prime}_{{\mathcal Ad}}}+ 4{\mathcal S_{{\mathcal P}}} + 2 {\mathcal S_{{\mathcal G}}} + 2 {\mathcal S_{{\mathcal C}}} 
\end{equation} 
\noindent
where ${\mathcal S^{\prime}_{{\mathcal Ad}}}$ = $\sum_{i=1}^{n}2(n-1)$, is an additional column required to perform an in-place circuit.
 
It is also useful to calculate the optimal in-place MBQCLA circuit resources. According to Equation [\ref{opteqgen}],
\begin{equation}
{\mathcal S}_{optimal} = {\mathcal X}_{TPG}{\mathcal R}_{TPG}+{\mathcal X}_{CNOT}{\mathcal R}_{CNOT}+{\mathcal X}_{NOT}{\mathcal R}_{NOT}.
\end{equation}
\noindent
By the use of Table [\ref{table0a}], one can obtain
\begin{equation}
\label{newopt}
{\mathcal S}_{optimal} =162(w(n)+\lfloor log_{2}(n-1) \rfloor+\lfloor log_{2}(n) \rfloor-w(n-1))+542n-395.
\end{equation} 

\subsection[MBQCLA Latencies]{\label{sec:level3g}MBQCLA Latencies} 

As discussed in section \ref{sec:level3p}, a carry-lookahead addition in MBQC can reach $O(log~n)$ time due to the constant scale depth of primitive gates in MBQC. The depth of a Toffoli gate in MBQC is 2, and our circuit does not change the original behavior. In addition to the Toffoli-dependent rounds, the QCLA requires a small number rounds of CNOTs and NOTs, each of which adds one to the circuit depth, giving a total of out-of-place and in-place MBQCLA depths are $2(\lfloor log_{2}(n)\rfloor + \lfloor log_{2}\frac{2n}{3}\rfloor )+13$ and 2$\left(\lfloor log_{2}(n)\rfloor+\lfloor log_{2}(n-1)\rfloor+\lfloor log_{2}\frac{n}{3}\rfloor+\lfloor {log_{2}}\frac{n-1}{3}\rfloor+14\right)$, respectively.

\subsection[Optimized In-place MBQCLA]{\label{sec:level3d}Optimized In-place MBQCLA}

We can optimize the MBQCLA spatial resources in several ways. First, by relaxing the Manhattan constraints on the physical geometry, we can use a graph state, which requires fewer communication resources if it can be physically implemented. The graph state adder will be presented in section \ref{sec:3ef}. 

In this section, we retain the Manhattan constraint but optimize the circuit. The idea of the bent network\cite{Raussendorf03}, imagining that logical qubits are propagated through traces on a single-layer two-dimensional surface, is used to reduce the horizontal resources of MBQCLA.

\subsubsection[Bent Network in Quantum Carry-Lookahead Adder]{\label{sec:level3e}Bent Network in Quantum Carry-Lookahead Adder}
\noindent 

Contrary to the usual quantum circuit assumption that the horizontal axis relates to logical time, in a "bent'' network the temporal axis flows freely in the spatial layout. The consequence is that a more compact circuit can be constructed.

\begin{figure} 
\centerline{\psfig{file=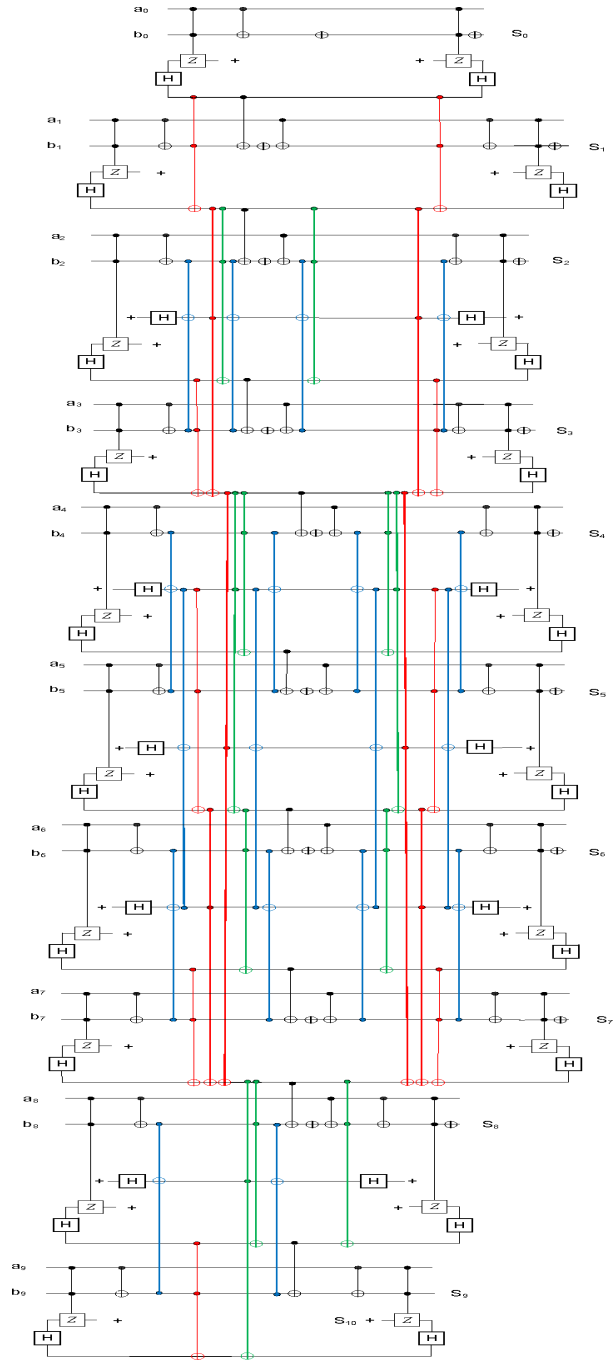, width=80mm, height=180mm}} 
\vspace*{13pt}
\caption{\label{motion1x}Optimized in-place circuit. The low $n$-1 bits of the carry string output are tucked into the interior of the circuit by bending the network.}
\end{figure}

If we apply this bent network method to MBQCLA, we also find that we can reduce the horizontal size of the circuit. The bent form of VBE is purely rectangular, but MBQCLA is not as regular. The horizontal size for every logical qubit position will depend on the number of quantum gates, since it will vary along the register as shown in Fig.\ref{motion1x}.  The illustration of a bent network implementation for $n$=10 is given in Fig.\ref{optimizedin-placembqc}.

\subsubsection[Optimized Circuit Formulation]{\label{sec:level3f}Optimized Circuit Formulation}

\begin{figure} 
\centerline{\psfig{file=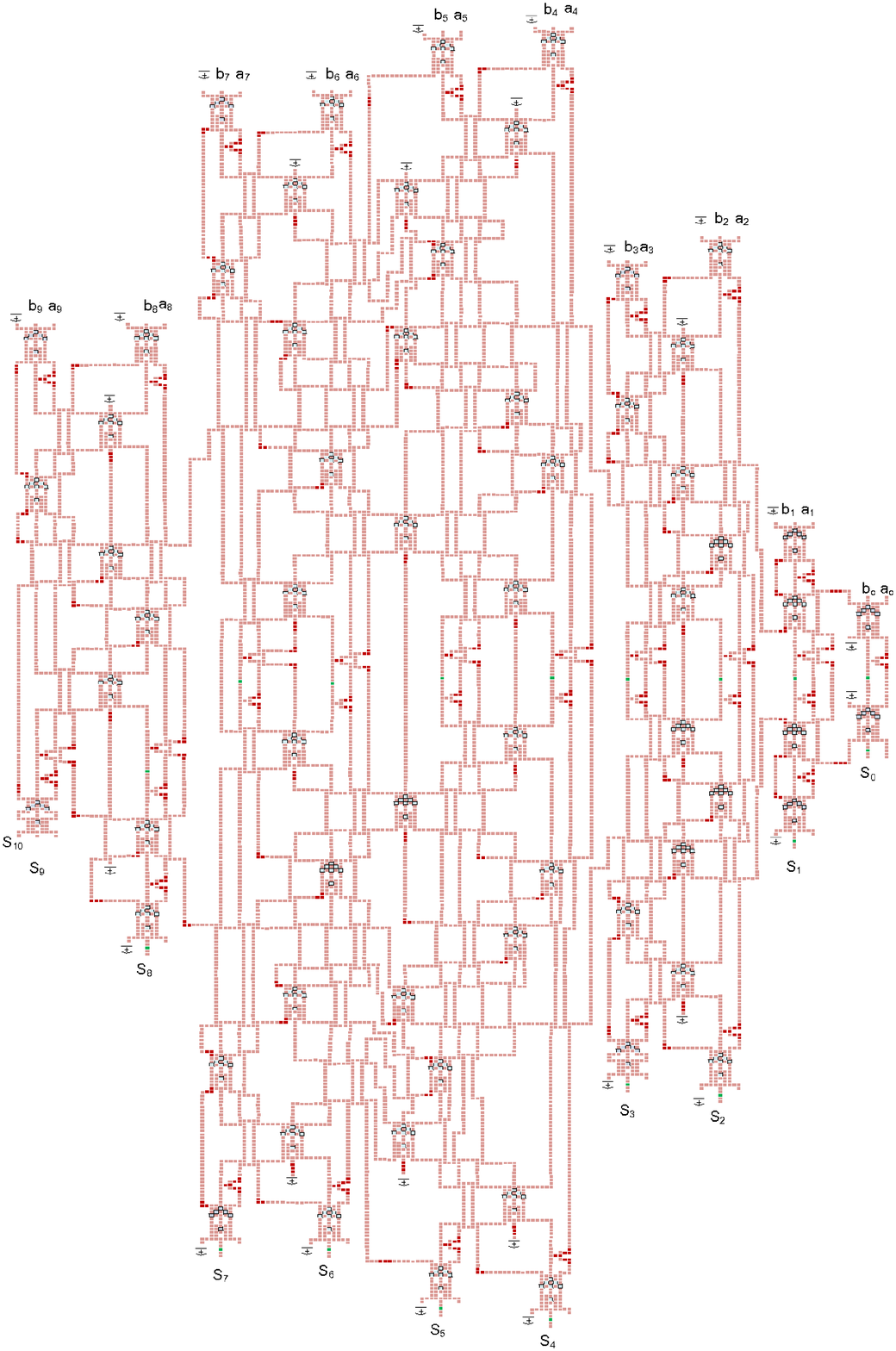, width=120mm, height=180mm}} 
\vspace*{13pt}
\caption{\label{optimizedin-placembqc}The optimized in-place MBQCLA circuit forms a diamond-like circuit. Hand optimization of the circuit reduced the size by $\approx$ 12 $\%$.}
\end{figure}

To optimize the circuit, we take small groups of qubits, or subregisters, and slide them toward the middle of the circuit. As can be seen in Fig.\ref{optimizedin-placembqc}, bending the network  can reduce the cluster resources required near qubits $a_{0}$, $a_{1}$, $a_{2}$, $a_{3}$, $a_{4}$, $a_{6}$, $a_{8}$, and $a_{9}$ by the amount: 

\begin{equation}
\sum_{\it{i} = 0}^{\it{n-1}}({\mathcal C}_{i}{\mathcal A}_{i}{\mathcal W})
\end{equation} 
\noindent
where \textit{n} = number of logical qubits, ${\mathcal C}_{i}$= the number of rounds (columns) that $i$th-subregister moves, ${\mathcal A}_{i}$= number of logical qubits in the $i$th-subregister (usually 3, sometimes 4), and ${\mathcal W}$= width of Toffoli Phase Gate. 
\noindent
For $n$=10, this manual optimization of horizontal communication results in a reduction of $\approx$ 12 ${\mathcal \%}$ for spatial resources, or $\approx$ 3822 qubits.

\subsection[GSQCLA]{\label{sec:3ef}Graph State Quantum Carry-Lookahead Adder (GSQCLA)}

\begin{figure} 
\centerline{\psfig{file=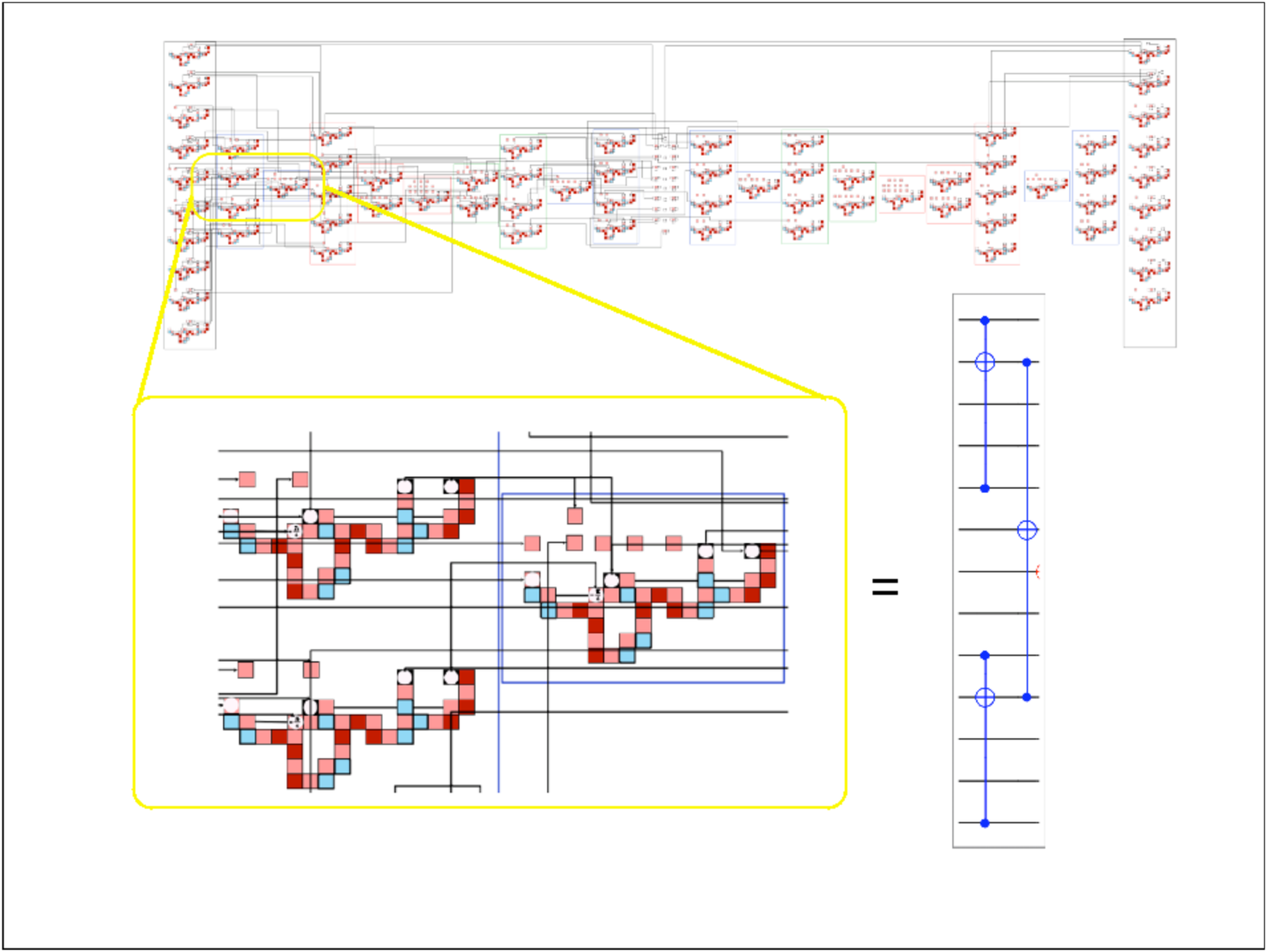, width=120mm, height=80mm}} 
\vspace*{13pt}
\caption{\label{GSQCLA2}MBQCLA figure allowing the use graph for entanglement and communication.}
\end{figure}

In the previous sections, we presented cluster state adders. Here, we present a graph state adder. The GSQCLA is simpler and follows more directly from the original QCLA definition. For MBQC in graph states, we assume that the restriction to Manhattan physical geometry is lifted, and arbitrary entanglement operations between qubits are allowed.  The vertices follow the graphical notation of MBQC in cluster states but with the three additional types of vertices:$\{$measured input qubits in an arbitrary angle, $-\frac{\pi}{4}$, and $\frac{\pi}{4}$$\}$ and there are two types of edges : entanglement and input/output information flows. We choose ${\mathcal {CNOT}}^{<4>}$ and ${\mathcal {CCNOT}}^{<39>}$ for running QCLA. When two GSQC quantum gates are concatenated, the output qubits of one become the input of the other. The birds eye view of the in-place GSQCLA is given in Figure \ref{GSQCLA2}.

The circuit depth of graph state is identical to that of MBQC, so we focus here on the number of qubits and entanglement operations.

Concatenating quantum gates in graph states can reduce the number of qubits whilst the number of entanglement operations is invariant\cite{Voufo1}. The number of entanglement operations in GSQCLA is

\begin{equation}
\label{entops}
\sum_{k} E_{k} {\mathcal X_{k}} 
\end{equation}

\noindent
Where  $k~\epsilon$ $\{$Toffoli Phase, CNOT, and NOT Gates$\}$, $E_{k}$ = the number of entanglement operations in type $k$ gates, and ${\mathcal X_{k}}$ = the number of gates of type $k$. 

The number of qubit resources before the removal of unnecessary measurement is given by Eq.\ref{opteqgen}. After the adjustment, the qubits resources is

\begin{equation}
\label{opteqgen2}
\sum_{m}{\mathcal X_{i}}{\mathcal R_{i}}-\sum_{l}{\mathcal N_{l}}({\mathcal Q_{P}}, {\mathcal Q_{G}}, {\mathcal Q_{C}}, {\mathcal Q_{add}})
\end{equation}  

\noindent
Where  $i~\epsilon$ $\{$Toffoli Phase, CNOT, and NOT Gates$\}$, ${\mathcal N_{l}}$ = number of removed qubits of type $l$ circuit and ${\mathcal Q_{P}}=\sum_{t_{p}=1}^{\lfloor log(n)\rfloor-1}2(\lfloor \frac{n}{2^{t_{p}}} \rfloor-1),~{\mathcal Q_{G}}=\sum_{t_{g}=1}^{\lfloor log(n)\rfloor}3\lfloor \frac{n}{2^{t_{g}}} \rfloor,~{\mathcal Q_{C}}=\sum_{t_{c}=1}^{\lfloor log_{2}(\frac{2n}{3})\rfloor-1}3\lfloor \frac{(n-2^{t_{c}-1})}{2^{t_{c}}} \rfloor,$ and ${\mathcal Q_{add}}$ are the number of removed qubits in $P$, $G$, $C$, and additional rounds in the circuit, respectively. The formulation of ${\mathcal Q_{add}}$ varies depending on the type of circuit.

We know from Table (\ref{table0a}) that QCLA requires 10$n$-3$w(n)$-3$w(n-1)$-3$\lfloor log_{2}(n) \rfloor$-3$\lfloor log_{2}(n-1) \rfloor$-7 Toffoli Phase Gates and that the Toffoli Phase Gate uses 39 qubits and 43 entangling operations. Therefore, based on the above formulations, the entanglement operations for the out-of-place GSQCLA is
\begin{equation}
\label{cct1}
224n-129\left(w(n)-\lfloor log_{2}(n)\rfloor\right)-46.
\end{equation}
\noindent
The number of qubits for this circuit is
 \begin{eqnarray}
\label{cct2}
201n-117\left(w(n)-\lfloor log_{2}(n)\rfloor\right)-2\sum_{t_{p}=1}^{\lfloor log_{2}(n)\rfloor-1}2(\lfloor \frac{n}{2^{t_{p}}} \rfloor-1)\nonumber
\\-\sum_{t_{g}=1}^{\lfloor log_{2}(n)\rfloor}3\lfloor \frac{n}{2^{t_{g}}} \rfloor-\sum_{t_{c}=1}^{\lfloor log_{2}(\frac{2n}{3})\rfloor-1}3\lfloor \frac{(n-2^{t_{c}-1})}{2^{t_{c}}} \rfloor-43,
\end{eqnarray}
\noindent
or roughly $201n$ for large $n$. This formulation is obtained after concatenating the GSQCLA quantum gates and adjusting their qubits resources to form the circuit. Similarly, the in-place GSQCLA has
\begin{equation}
\label{ipgs}
444n-129\left(w(n)-w(n-1)-\lfloor log_{2}(n)\rfloor-\lfloor log_{2}(n-1)\rfloor\right)-318
\end{equation}
\noindent
entanglement operations. The number of qubits for the in-place GSQCLA is
\begin{eqnarray}
\label{ipgsqla}
410n-117\left(w(n)-w(n-1)-\lfloor log_{2}(n)\rfloor-\lfloor log_{2}(n-1)\rfloor\right)-4\sum_{t_{p}=1}^{\lfloor log_{2}(n)\rfloor-1}2(\lfloor \frac{n}{2^{t_{p}}} \rfloor-1) \nonumber
\\-2\sum_{t_{g}=1}^{\lfloor log_{2}(n)\rfloor}3\lfloor \frac{n}{2^{t_{g}}} \rfloor-2\sum_{t_{c}=1}^{\lfloor log_{2}(\frac{2n}{3})\rfloor-1}3\lfloor \frac{(n-2^{t_{c}-1})}{2^{t_{c}}} \rfloor-261,~~~~~~~~~~~~~~~~
\end{eqnarray}
\noindent
about twice the size of the out-of-place version.

\subsection[Resource Comparison]{\label{sec:level4dc}Resource Comparison}

\begin{figure}[!ht]
\centerline{\psfig{file=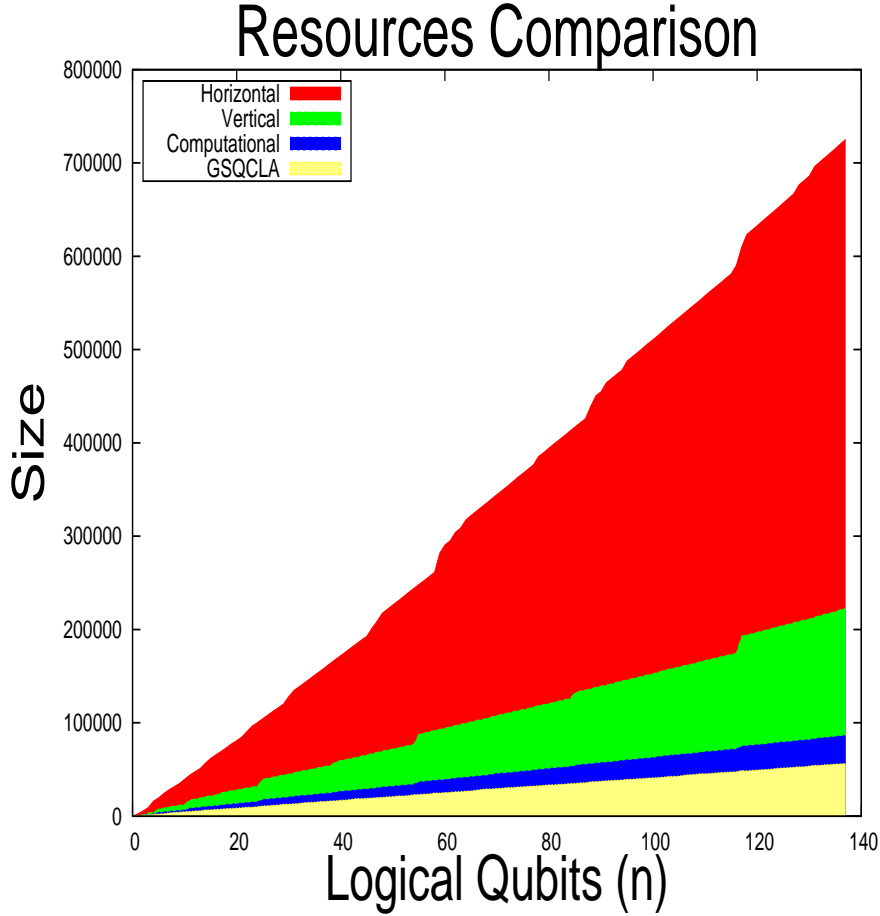, width=120mm, height=120mm}} 
\vspace*{13pt}
\caption{\label{comparisonresources}Comparison of computational and communication resources in in-place MBQCLA circuit. The bottom line on the graph represents the ideal circumstance for the computational resource and the other two show the circuit with additional resources for the horizontal and the vertical communications.}
\end{figure}

Figure \ref{comparisonresources} plots the resources required for the in-place MBQCLA, as derived in Equations (\ref{optinpl}), (\ref{newopt}), and (\ref{ipsize}). The red area, which represents the horizontal communication costs of MBQCLA, is $\approx$ 77  ${\mathcal \%}$ of the qubits in the cluster. The light green area, showing the costs of MBQCLA circuit vertical communication, consumes $\approx$ 11  ${\mathcal \%}$.  The cost of the computational circuit shown by the light blue area is $\approx$ 12  ${\mathcal \%}$ of the spatial resources. The light yellow area represents  the qubits resources for the in-place GSQCLA which costs $\approx$ 9 $\%$ in spatial resources. 

\pagebreak
\section[Conclusion]{\label{sec:level4}Conclusion}
\noindent
In this work, the circuit designs for several forms of a measurement-based quantum carry-lookahead adder (MBQCLA) and graph-state quantum carry-lookahead adder (GSQCLA) are presented. We have shown the resources required to perform the quantum carry-lookahead adder in cluster state as a function of the number of logical qubits, width of quantum gates, and number of qubits in quantum gates. By bending the network and removing the border between the rounds, the optimization of the in-place MBQCLA circuit changes its shape from a rectangle to a diamond-like form. The proposed evaluation methods for the cost and performance of application circuits for MBQC will be useful for large scale quantum computer architecture, since application circuits for quantum computers will need optimization similar to that done for classical computer technology. This work has shown the value of finding application algorithms that match the strengths of measurement-based quantum computation.  

\vspace*{20pt}
\paragraph[Acknowledgments]{\label{sec:5}Acknowledgments}

We would like to thank Seth Lloyd, Kohei M. Itoh, Austin Fowler, and Achmad Husni Thamrin for fruitful discussions. This work was supported in part by Grant-in-Aid for Scientific Research by MEXT, Specially Promoted Research No. 18001002 and in part by Special Coordination Funds for Promoting Science and Technology.

\begin{figure} 
\centerline{\psfig{file=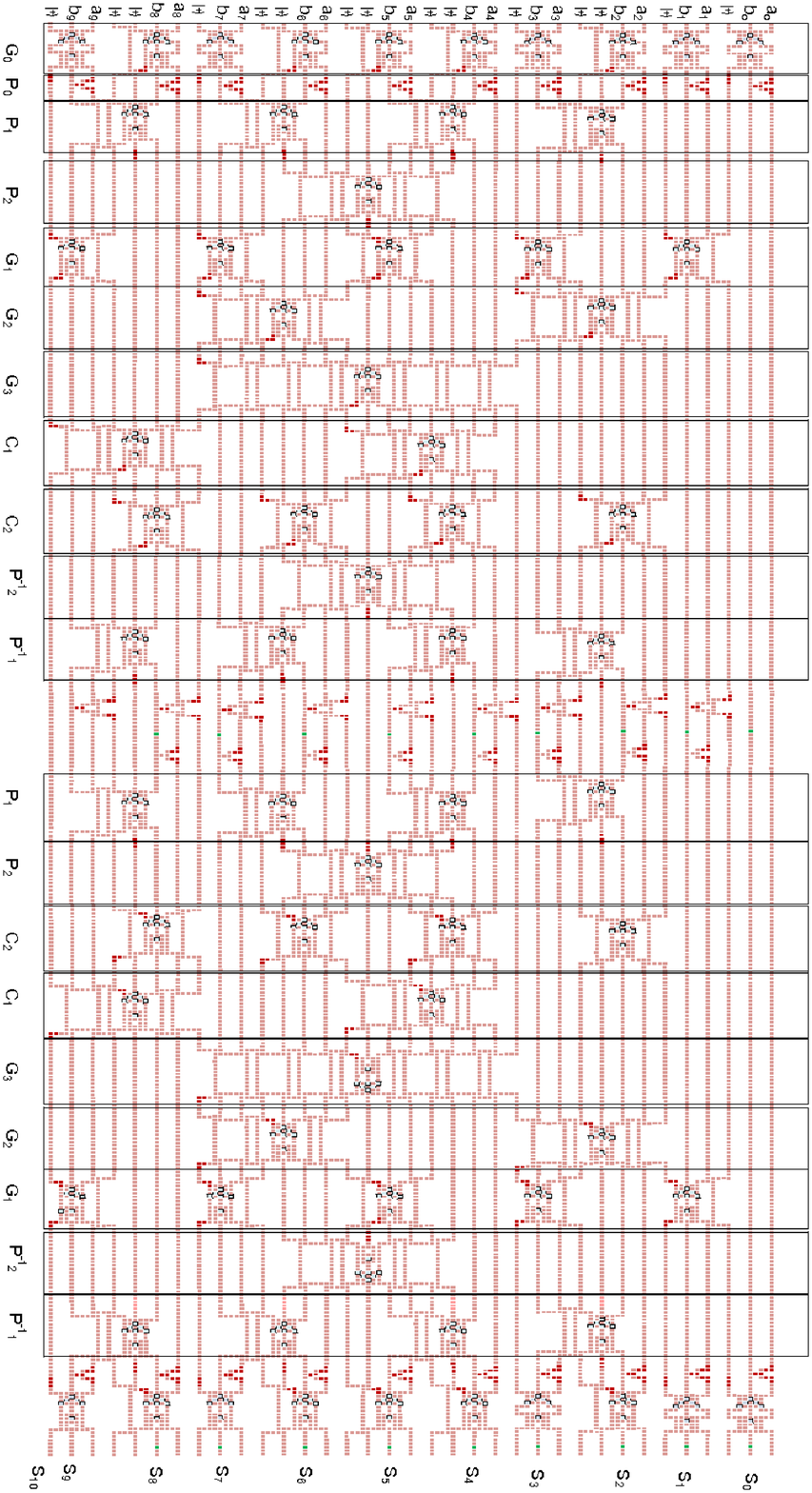, width=120mm, height=180mm}} 
\vspace*{13pt}
\caption{\label{mbqc-in-place}In-place MBQCLA. For $n$=10, the circuit consists of:8 addition circuits, 18 carry networks (4 Propagate, 6 Generate, 4 Inverse Propagate and 4 Carry networks) 
}
\end{figure}

\pagebreak
\section*{\label{appA}  Appendix A MBQC Gates and Graphical Notation}

\noindent

\begin{figure} [!htp]
\centerline{\psfig{file=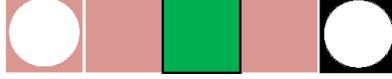, width=5.2cm}} 
\vspace*{13pt}
\caption{\label{NOTgateMBQC} NOT gate in MBQC. The input qubit and all pink qubits are measured in the $\sigma_{x}$-eigenbasis and the framed-green qubit is measured in an adaptive basis depending on the measurement outcome of qubit 2. Rotation measurement operator on $xy$-plane around $z$-axis with angle $\pi$ on qubit 3 makes the qubit 3 not adaptive.}
\end{figure}

To illustrate MBQC, we detail the operation of the NOT gate, as shown in Figure \ref{NOTgateMBQC}. The cluster contains 5 qubits where ${\mathcal C}_{input}$ is qubit 1, ${\mathcal C}_{machine}$ is qubits 2, 3 and 4 and  ${\mathcal C}_{output}$ is qubit 5. We begin from the cluster state eigenvalue equations for 5 qubits: 

\begin{equation}
\sigma^{(1)}_{x}\sigma^{(2)}_{z}\left|\phi\right\rangle_{{\mathcal C}(\textbf{NOT})}=\left|\phi\right\rangle_{{\mathcal C}(\textbf{NOT})}
\end{equation}

\begin{equation}
\sigma^{(1)}_{z}\sigma^{(2)}_{x}\sigma^{(3)}_{z}\left|\phi\right\rangle_{{\mathcal C}(\textbf{NOT})}=\left|\phi\right\rangle_{{\mathcal C}(\textbf{NOT})}
\end{equation}

\begin{equation}
\sigma^{(1)}_{z}\sigma^{(2)}_{z}\sigma^{(3)}_{x}\sigma^{(4)}_{z}\left|\phi\right\rangle_{{\mathcal C}(\textbf{NOT})}=\left|\phi\right\rangle_{{\mathcal C}(\textbf{NOT})}
\end{equation}

\begin{equation}
\sigma^{(3)}_{z}\sigma^{(4)}_{x}\sigma^{(5)}_{z}\left|\phi\right\rangle_{{\mathcal C}(\textbf{NOT})}=\left|\phi\right\rangle_{{\mathcal C}(\textbf{NOT})}
\end{equation}

\begin{equation}
\sigma^{(1)}_{z}\sigma^{(4)}_{z}\sigma^{(5)}_{x}\left|\phi\right\rangle_{{\mathcal C}(\textbf{NOT})}=\left|\phi\right\rangle_{{\mathcal C}(\textbf{NOT})}
\end{equation}

After obtaining the quantum correlation of the cluster state, two measurement steps are performed: $first$, the $\sigma_{x}$ measurement on qubit 2 and qubit 4,
\begin{equation} 
\left|\phi\acute{}\right\rangle_{{\mathcal C}(\textbf{NOT})}=
{\mathcal P}^{(2)}_{x,s_{2}}{\mathcal P}^{(4)}_{x,s_{4}}\left|\phi\right\rangle_{{\mathcal C}(\textbf{NOT})}\end{equation}
\noindent
This first measurement converts the initial quantum correlation to the eigenvalue equations:

\begin{equation}
\label{a6}
\sigma^{(1)}_{x}\sigma^{(3)}_{x}\sigma^{(5)}_{x}\left|\phi\acute{}\right\rangle_{{\mathcal C}(\textbf{NOT})} = \left|\phi\acute{}\right\rangle_{{\mathcal C}(\textbf{NOT})}
\end{equation}

\begin{equation}
\label{a7}
\sigma^{(1)}_{z}\sigma^{(3)}_{z}\left|\phi\acute{}\right\rangle_{{\mathcal C}(\textbf{NOT})} = (-1)^{s_{2}} \left|\phi\acute{}\right\rangle_{{\mathcal C}(\textbf{NOT})}
\end{equation}

\begin{equation}
\label{a8}
\sigma^{(3)}_{z}\sigma^{(5)}_{z}\left|\phi\acute{}\right\rangle_{{\mathcal C}(\textbf{NOT})} = (-1)^{s_{4}} \left|\phi\acute{}\right\rangle_{{\mathcal C}(\textbf{NOT})}
\end{equation}

Furthermore, the eigenbasis of $\vec{r}_{xy}((-1)^{s_{2}}(-\eta)).\vec{\sigma}$, where $\vec{r}_{xy}.\vec{\sigma}$= cos($\eta$) $\sigma_{x}$+sin($\eta$) $\sigma_{y}$, is chosen as the measurement basis on qubit 3 to realize the operation $\textbf{NOT}$ by measurement pattern ${\mathcal M}$($\textbf{NOT}$). Mathematically, it can be expressed by: 

\begin{equation}\left|\psi\right\rangle_{{\mathcal C}(\textbf{NOT})}=
{\mathcal P}^{(3)}_{xy(\eta)}\left|\phi\acute{}\right\rangle_{{\mathcal C}(\textbf{NOT})},\end{equation}
\noindent
where ${\mathcal P}^{(3)}_{xy(\eta)}$=$\frac{1+(-1)^{s_{3}}\vec{r}_{k}.\vec{\sigma}^{(k)}}{2}$. This second measurement generates two eigenvalue equations from equation (\ref{a6}), (\ref{a7}) and (\ref{a8}), which obey $Theorem~1$ of Raussendorf et. al: 

\begin{equation}
\sigma^{(1)}_{x}U^{(5)}[-\eta]\sigma^{(5)}_{x}U^{(5)\dagger}[\eta]\left|\psi\right\rangle_{{\mathcal C}(\textbf{NOT})} = \left|\psi\right\rangle_{{\mathcal C}(\textbf{NOT})}
\end{equation}

\begin{equation}
\sigma^{(1)}_{z}U^{(5)}[-\eta]\sigma^{(5)}_{z}U^{(5)\dagger}[\eta]\left|\psi\right\rangle_{{\mathcal C}(\textbf{NOT})} = (-1)^{s_{2}+s_{4}}\left|\psi\right\rangle_{{\mathcal C}(\textbf{NOT})}
\end{equation}
\noindent
By choosing $\eta$ = $\pi$, these equations give a NOT-gate. This method can be broadened to perform quantum gates on a large-scale cluster state system. Similar to the work of Leung\cite{Debbie01}, quantum computation can be achieved in cluster states depending on choice of measurement patterns.

\begin{figure} [!htp]
\centerline{\psfig{file=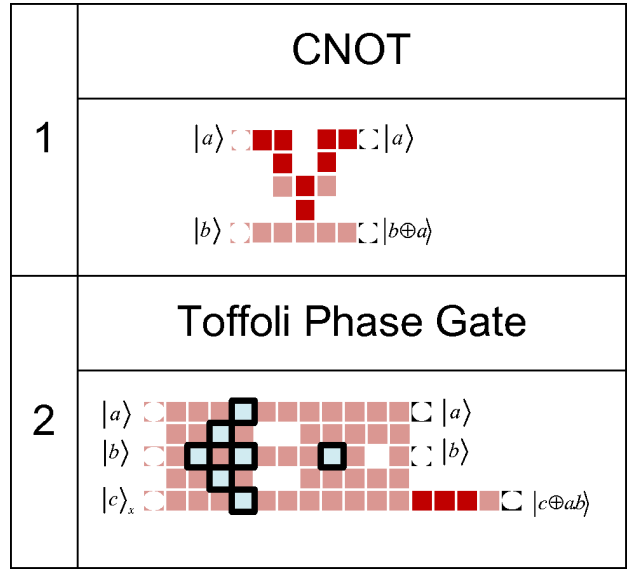, width=8.2cm}} 
\vspace*{20pt}
\caption{\label{mbqcquantumgates}Quantum gates in Measurement-Based Quantum Computation. Due to the relationship between CCNOT and Toffoli Phase Gate (TPG), CCNOT=${H_{t}}$$({TPG})$${H_{t}^{\dagger}}$, the target qubit can be chosen arbitrarily by putting a Hadamard gate on the chosen qubit.}
\end{figure}

\begin{figure} 
\vspace*{13pt}
\centerline{\psfig{file=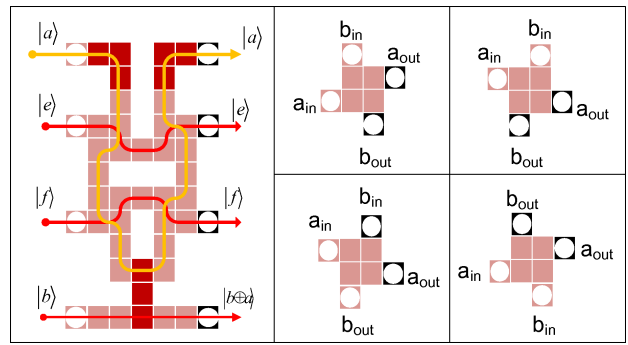, width=8.2cm}} 
\vspace*{20pt}
\caption{\label{Nonadjacentcomputation} Non-adjacent computation. The implementation of four type SWAP gates to propagate the information up-to-down
of non-adjacent qubits for performing CNOT. Qubit $|$$a$$>$ acts as the control qubit and $|$$b$$>$ is the target qubit.}
\end{figure}

\noindent
Generally, every quantum gate contains $C_{\it{l}}$ lattice qubits with $m$ measurements, $C_{\it{w}}$ width, and $C_{\it{h}}$ height. In this paper, we will use Raussendorf \textit{et al.}`s quantum gates model\cite{Raussendorf03}. As shown in Figure \ref{mbqcquantumgates} and Figure \ref{Nonadjacentcomputation}, CNOT and the Toffoli Phase gate can be performed using measurement on 15 and 54 cluster qubits, respectively.
\section*{\label{proce12} Appendix B Out-of-place and In-place Procedures for Abstract Quantum Carry-Lookahead Adder}

\begin{figure} [!htp]
\centerline{\psfig{file=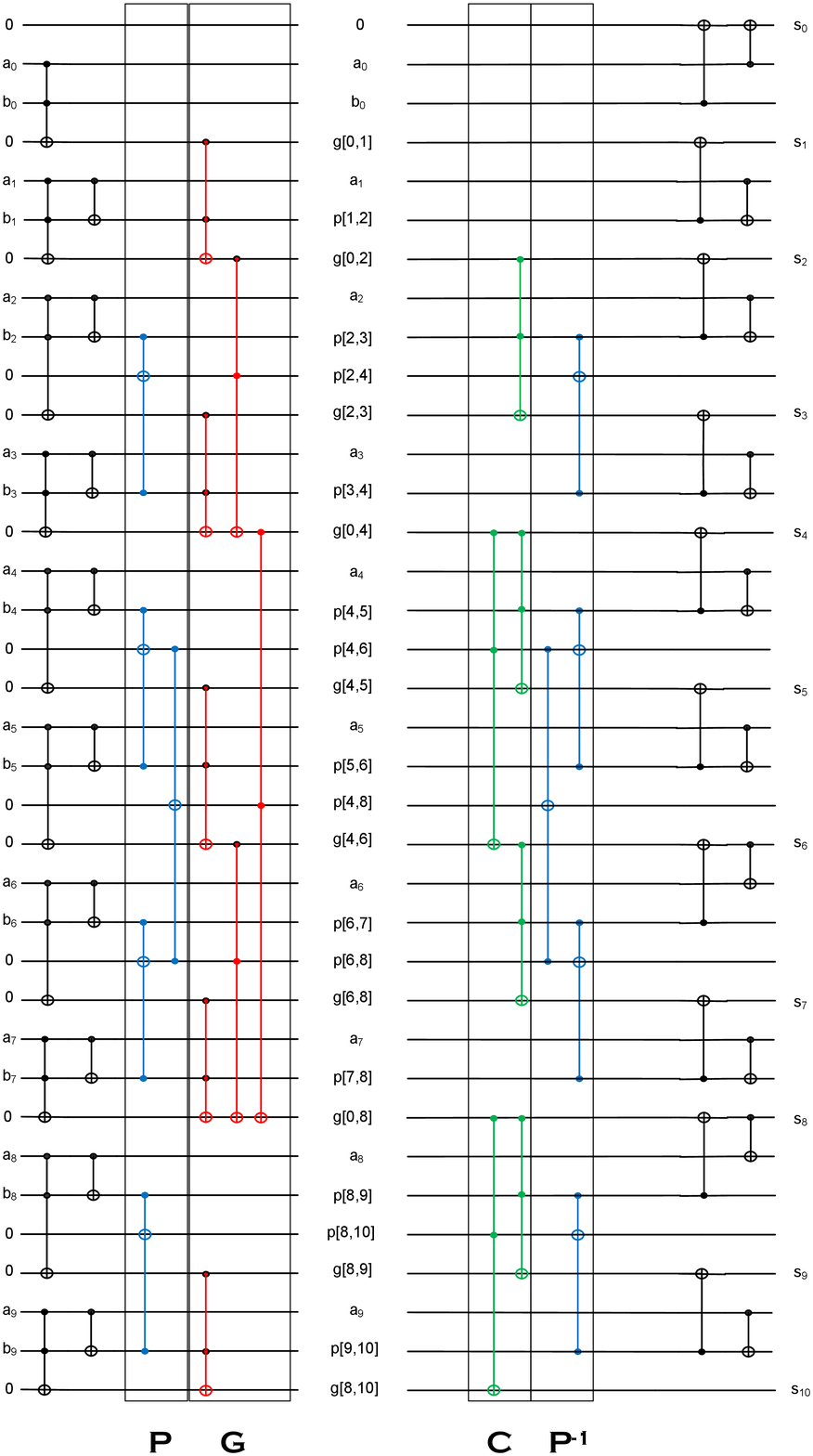, width=100mm, height=180mm}} 
\vspace*{13pt}
\caption{\label{abstractqcla}Abstract out-of-place Quantum Carry-Lookahead Adder for $n$=10. The blue lines are the Propagate and Inverse Propagate networks; the red line is the Generate network and the green line is the Carry network. The low $n$-1 bits of the carry string are not yet erased .}
\end{figure}

Here we summarize the QCLA circuits as proposed by Draper \textit{et al.}\cite{Draper06}. The circuit for out-of-place addition as shown in Figure \ref{abstractqcla} has the form:
\begin{enumerate}
	\item For 0$\leq$$i$$<$$n$, $Z$[$i$+1]$\oplus$=$A$[$i$]$B$[$i$] setting $Z$[$i$+1]=$g$[$i$, $i$+1].
	\item For 0$\leq$$i$$<$$n$, $B$[$i$]$\oplus$=$A$[$i$] setting $B$[$i$]=$p$[$i$, $i$+1] for $i$$>$0 needed to run out-of-place addition circuit.
	\item Run the circuit of the P,G, and C networks. Upon completion, $Z$[$i$]=$c$$_{i}$ for $\geq$1.
	\item For 1$\leq$$\it{i}$$<$$n$, $Z$[$i$]$\oplus$=$B$[$i$]. Now for $i$$>$0, $Z$[$i$]=$a_{\textit{i}}$$\oplus$$b_{\textit{i}}$$\oplus$$c_{\textit{i}}$=$s_{i}$. For $i$=0, $Z$[$i$]=$b_{\textit{i}}$. 
	\item Set $Z$[0]$\oplus$=$A$[$i$]. For 1$\leq$$i$$<$$n$, $B$[$i$]$\oplus$=$A$[$i$]. This fixes $Z$[0], and resets $B$ to initial value.   
\end{enumerate}
\noindent
The addition circuit for in-place operation has form:
\begin{enumerate}
	\item For 0$\leq$$i$$<$$n$, $Z$[$i$+1]$\oplus$=$A$[$i$]$B$[$i$] setting $Z$[$i$+1]=$g$[$i$, $i$+1].
	\item For 0$\leq$$i$$<$$n$, $B$[$i$]$\oplus$=$A$[$i$] setting $B$[$i$]=$p$[$i$, $i$+1] for $i$$>$0 and $B$[$i$]=$s$$_{0}$.
	\item Run the circuit of the P,G, and C networks. Upon completion, $Z$[$i$]=$c$$_{i}$ for $\geq$1.
	\item For 1$\leq$$i$$<$$n$, $B$[$i$]$\oplus$=$Z$[$i$]. Now $B$[$i$]=$s$$_{i}$.
	\item For 0$\leq$$i$$<$$n$-1, $\neg$$B$ contains $\it{s}'$.
	\item For 1$\leq$$i$$<$$n$-1, $B$[$i$]$\oplus$=$A$[$i$].
	\item Run the P, G, and network in reverse. Upon completion, $Z$[$i$+1]=$a_{\it{i}}$$s'_{\it{i}}$ for 0$\leq$$i$$<$$n$-1, and $B$=$a_{\it{i}}$$\oplus$$\it{s}'_{\it{i}}$ for 1$\leq$$i$$<$$n$.
	\item For 1$\leq$$i$$<$$n$-1, $B$[$i$]$\oplus$=$A$[$i$].
	\item 0$\leq$$i$$<$$n$-1, $Z$[$i$+1]$\oplus$=$A$[$i$]$B$[$i$].
	\item 0$\leq$$i$$<$$n$-1, $\neg$$B$.     
\end{enumerate}

\noindent
The resources for each quantum gates in the abstract in-place QCLA circuit is provided in the below table:

\begin{table}[!htp]
\tbl{\label{table0a}Logic Gates Resources in Abstract In-Place QCLA}
{\begin{tabular}{||l|lr||}\hline
\textbf{Quantum Gate}  & \multicolumn{2}{|c||}{\textbf{Resource}}\\ \hline
NOT & 2$n$-2 &\\ \hline 
CNOT  & 4$n$-5 &\\  \hline
TPG  & 10$n$-3$w(n)$-3$w(n-1)$-3$\lfloor log(n) \rfloor$-3$\lfloor log_{2}(n-1) \rfloor$-7 &\\ \hline
\end{tabular}}
\end{table}
\pagebreak
\section*{\label{appc}  Appendix C Requirements and Performance for Out-of-place and In-place MBQCLA}

\begin{table}[!htp]

\tbl{\label{table1}Requirements and Performance of the out-of-place MBQCLA}
{\begin{tabular}{||l|lr||}\hline
\textbf{Parameter}  & \multicolumn{2}{|c||}{\textbf{Value}}\\ \hline
Pitch & 4 &\\ \hline 
Variables  & ${\mathcal V}(n)$ = 4$\it{n}$-$\lfloor$log$_{2}$$\it{n}$$\rfloor$+1 &\\ 
(Logical& &\\ 
Qubits)& &\\ \hline
Width & ${\mathcal Width}(n)$=15$\times$($\lfloor$log$_{2}$($\it{n}$)$\rfloor$+$\lfloor$log$_{2}$($\it{n}$/3)$\rfloor$)+85&\\ \hline
Height & ${\mathcal Height}(n)$=4$\times$(4$\it{n}$-$\it{w(n)}$-$\lfloor$log$_{2}$$\it{n}$$\rfloor$+1)-3&\\ \hline
Area & ${\mathcal Height}(n)$$\times$${\mathcal Width}(n)$ = &\\ 
& (4$\times$(4$\it{n}$-$\it{w(n)}$-$\lfloor$log$_{2}$$\it{n}$$\rfloor$+1)-3)$\times$ &\\ 
& (14$\times$($\lfloor$log$_{2}$($\it{n}$)$\rfloor$+$\lfloor$log$_{2}$($\it{n}$/3)$\rfloor$)+85) &\\ \hline
Number of & (4$\it{n}$-$\it{w(n)}$-$\lfloor$log$_{2}$$\it{n}$$\rfloor$+1)$\times$ &\\
Clustering & (15$\times$$\lfloor$log$_{2}$($\it{n}$)$\rfloor$+$\lfloor$log$_{2}$($\it{n}$/3)$\rfloor$+85)-1)+ &\\
Operations & (15$\times$$\lfloor$log$_{2}$($\it{n}$)$\rfloor$+$\lfloor$log$_{2}$($\it{n}$/3)$\rfloor$+85))$\times$ &\\
& (4$\it{n}$-$\it{w(n)}$-$\lfloor$log$_{2}$$\it{n}$$\rfloor$)&\\ \hline
Circuit & $\lfloor$log$_{2}$($\it{n}$)$\rfloor$+$\lfloor$log$_{2}$($\it{n}$/3)$\rfloor$+7&\\
Depth & &\\ \hline
Size& -3271+899$n$-419$w(n)$-377$\lfloor$$log_{2}$$(n)$$\rfloor$+56$n$$\lfloor$$log_{2}$($2n$/3)$\rfloor$-14$w(n)$$\lfloor$$log_{2}$($2n$/3)$\rfloor$&\\
(Number of& +42$n$$\lfloor$$log_{2}$$(n)$$\rfloor$+168$n$$\lfloor$$log_{2}$$(n)$$\rfloor$-14$\lfloor$$log_{2}$($2n$/3)$\rfloor$$\lfloor$$log_{2}$$(n)$$\rfloor$-42$\lfloor$$log_{2}$$(n)$$\rfloor$$^{2}$&\\ 
Qubits)&+6$\times$(2\textit{n}+$\sum_{t_{p}=1}^{log_{2}(n-1)}$ 2($n$-2$^{t_{p}}$)+$\sum_{t_{g}=1}^{log_{2}(n)}$ 2(2($\lfloor log_{2}(n)\rfloor$)+ (2$^{t_{g}}$)($\lfloor log_{2}(t_{g})\rfloor$)&\\
&+$\sum _{t_{c}=1}^{\lfloor log_{2}\frac{2n}{3}\rfloor}$ 2($n$-$\lfloor log_{2}(t_{c})\rfloor$))&\\ \hline
\end{tabular}}
\end{table}

\begin{table}[!htp]
\vspace*{20pt}
\tbl{\label{table2}Requirements and Performance of the in-place MBQCLA}
{\begin{tabular}{||l|lr||}\hline
\textbf{Parameter}  & \multicolumn{2}{|c||}{\textbf{Value}}\\ \hline
Pitch & 4 &\\ \hline 
Variables  & ${\mathcal V}(n)$ = 4$\it{n}$-$\lfloor$$log_{2}$$\it{n}$$\rfloor$+1 &\\ 
(Logical& &\\ 
Qubits)& &\\ \hline
Width & ${\mathcal Width}(n)$=15$\times$($\lfloor$$log_{2}$$n$$\rfloor$+ $\lfloor$log$_{2}$$(n-1)$$\rfloor$+ $\lfloor$$log_{2}$$\frac{n}{3}$$\rfloor$+$\lfloor$log$_{2}$$\frac{n-1}{3}$$\rfloor$)+157&\\ \hline
Height & ${\mathcal Height}(n)$=4$\times$(4$\it{n}$-$\it{w(n)}$-$\lfloor$$log_{2}$$\it{n}$$\rfloor$+1)-3&\\ \hline
Area &${\mathcal Height}(n)$$\times$${\mathcal Width}(n)$ = &\\ 
& (4$\times$(4$\it{n}$-$\it{w(n)}$-$\lfloor$$log_{2}$$\it{n}$$\rfloor$+1)-3)$\times$ &\\ 
& 14$\times$($\lfloor$log$_{2}$$n$$\rfloor$+ $\lfloor$$log_{2}$$(n-1)$$\rfloor$+ $\lfloor$$log_{2}$$\frac{n}{3}$$\rfloor$+$\lfloor$$log_{2}$$\frac{n-1}{3}$$\rfloor$)+157 &\\ \hline
Number of & (4$\it{n}$-$\it{w(n)}$-$\lfloor$$log_{2}$$\it{n}$$\rfloor$+1)$\times$ &\\
Clustering & (15$\times$($\lfloor$log$_{2}$$n$$\rfloor$+ $\lfloor$$log_{2}$$(n-1)$$\rfloor$+ $\lfloor$$log_{2}$$\frac{n}{3}$$\rfloor$+$\lfloor$$log_{2}$$\frac{n-1}{3}$$\rfloor$)+156)+ &\\
Operations & (15$\times$($\lfloor$$log_{2}$$n$$\rfloor$+ $\lfloor$$log_{2}$$(n-1)$$\rfloor$+ $\lfloor$$log_{2}$$\frac{n}{3}$$\rfloor$+$\lfloor$$log_{2}$$\frac{n-1}{3}$$\rfloor$)+157)$\times$ &\\
& (4$\it{n}$-$\it{w(n)}$-$\lfloor$$log_{2}$$\it{n}$$\rfloor$)&\\ \hline
Circuit Depth &$\lfloor$$log_{2}$$n$$\rfloor$+ $\lfloor$log$_{2}$$(n-1)$$\rfloor$+ $\lfloor$$log_{2}$$\frac{n}{3}$$\rfloor$+$\lfloor$$log_{2}$$\frac{n-1}{3}$$\rfloor$+14&\\ \hline
Size& -3068+2896$n$-138$w(n-1)$-162$w(n)$+16$\lfloor$log$_{2}$$\frac{n-1}{3}$$\rfloor$&\\
&+64$n$$\lfloor$log$_{2}$$\frac{n-1}{3}$$\rfloor$-146$\lfloor$$log_{2}$($n$-1)$\rfloor$ &\\
(Number of&+64$n$$\lfloor$$log_{2}$($n$-1)$\rfloor$+16$\lfloor$$log_{2}$($n$/3)$\rfloor$+64$n$$\lfloor$$log_{2}$($n$/3)$\rfloor$&\\ 
&+21$\lfloor$$log_{2}$$(n)$$\rfloor$+64$n$$\lfloor$$log_{2}$$(n)$$\rfloor$ &\\ 
Qubits)& -16$\lfloor$$log_{2}$$(n-1)$$\rfloor$$\lfloor$$log_{2}$$(n)$$\rfloor$-16$\lfloor$$log_{2}$$\frac{n}{3}$$\rfloor$$\lfloor$$log_{2}$$(n)$$\rfloor$-16 $\lfloor$$log_{2}$$(n)$$\rfloor$$^{2}$+&\\
&6$\times$(4\textit{n}-1+ 4$\sum_{t_{p}=1}^{log_{2}(n-1)}$ 2($n$-2$^{t_{p}}$)+ 2$\sum_{t_{g}=1}^{log_{2}(n)}$ 2(2($\lfloor log_{2}(n)\rfloor$)+ (2$^{t_{g}}$)($\lfloor log_{2}(t_{g})\rfloor$)&\\
&+ 2$\sum _{t_{c}=1}^{\lfloor log_{2}\frac{2n}{3}\rfloor}$ 2($n$-$\lfloor logt_{c}\rfloor$)) &\\ \hline
\end{tabular}}
\end{table}

\end{document}